\documentclass[twocolumn,english,aps,superscriptaddres,prb,citeautoscripts]{revtex4-1}
\usepackage[T1]{fontenc}
\usepackage[latin9]{inputenc}
\usepackage{graphicx}
\usepackage{amsmath}
\usepackage{subscript}
\usepackage{babel}
\begin{document}

\title{Proton Irradiation-Decelerated Intergranular Corrosion of Ni-Cr Alloys in Molten Salt}
\author{W. Y. Zhou}
    \affiliation{Department of Nuclear Science and Engineering, Massachusetts Institute of Technology, Cambridge, USA}
\author{Y. Yang}
    \affiliation{National Center for Electron Microscopy, Molecular Foundry, Lawrence Berkeley National Laboratory, Berkeley, USA}
\author{G. Q. Zheng}
    \affiliation{Nuclear Reactor Laboratory, Massachusetts Institute of Technology, Cambridge, USA}
\author{K. B. Woller}
    \affiliation{Department of Nuclear Science and Engineering, Massachusetts Institute of Technology, Cambridge, USA}
\author{P. W. Stahle}
    \affiliation{Department of Nuclear Science and Engineering, Massachusetts Institute of Technology, Cambridge, USA}
\author{A. M. Minor}
    \affiliation{National Center for Electron Microscopy, Molecular Foundry, Lawrence Berkeley National Laboratory, Berkeley, USA}
    \affiliation{Department of Materials Science and Engineering, University of California, Berkeley, USA}
\author{M. P. Short}
    \email[Corresponding email address: ]{hereiam@mit.edu}
    \affiliation{Department of Nuclear Science and Engineering, Massachusetts Institute of Technology, Cambridge, USA}

\maketitle

The effects of ionizing radiation on materials often reduce to "bad news." Radiation damage usually leads to detrimental effects such as embrittlement \cite{odette1998recent}, accelerated creep \cite{bullough1980mechanisms}, phase instability \cite{russell1984phase}, and radiation-altered corrosion \cite{adamson2000effects}. This last point merits special attention. Elucidating synergies between radiation and corrosion has been one of the most challenging tasks impeding the deployment of advanced reactors \cite{was2019materials}, stemming from the combined effects of high temperature, corrosive coolants, and intense particle fluxes \cite{zinkle2009structural}. Here we report that proton irradiation significantly and repeatably decelerates intergranular corrosion of Ni-Cr alloys in molten fluoride salt at 650\textdegree C. We demonstrate this effect by showing that the depth of intergranular voids resulting from Cr leaching into the salt is reduced by the proton irradiation alone. Interstitial defects generated from proton irradiation result in radiation-enhanced diffusion, more rapidly replenishing corrosion-injected vacancies with alloy constituents, thus playing the crucial role in decelerating corrosion. Our results show that in industrially-relevant scenarios irradiation can have a positive impact, challenging our view that radiation damage always results in negative effects.

It has been well established that radiation accelerates corrosion of structural materials in today's reactors \cite{allen20125}, as in-core experiments \cite{asher1970effects} and accelerator studies \cite{wang2015oxidation} have shown. However, multiple competing mechanisms may be present \cite{cox2005some}. An overall increase in corrosion rate does not necessarily mean that each mechanism at play accelerates corrosion. The mechanisms and their relative strengths will necessarily change in different working fluids. In fact, observations of structural materials from the Molten Salt Reactor Experiment showed low corrosion rates \cite{IGNATIEV2012221}. This suggests that the influence of irradiation on corrosion in molten salts might be different from that in water-based systems. However, the reason(s) for this potential synergy have remained hidden for more than half a century.

Since corrosion products and normally-protective oxides are soluble in the salt, oxide-based passivation does not occur, making corrosion in molten salt different from aqueous corrosion \cite{sridharan2013corrosion}. The predominant mechanism of corrosion of Ni-based alloys in molten fluoride salts, such as the Ni-20Cr alloy in this study, is selective dissolution of Cr (here the most susceptible element) into the salt \cite{olson2009materials}. The net loss of atoms resulting from dissolution-based corrosion results in voids containing the salt \cite{ai2018influence}. The distribution of voids is often localized, preferentially occurring along grain boundaries \cite{yin2018effect}. One would therefore think that irradiation should increase the loss of Cr atoms by virtue of radiation enhanced diffusion of Cr towards the material/salt interface. Indeed Bakai et al. have observed a considerable increase in Ni-Mo corrosion in NaF-ZrF$_4$ salt under electron irradiation \cite{bakai2008combined}. However, we observed that proton irradiation actually decelerates corrosion in molten salt, contradicting radiation accelerated corrosion observed in water-based environments \cite{allen20125} and corroborating others \cite{hanbury2019oxide}. 

We have constructed a simultaneous irradiation/corrosion facility to explore how proton irradiation affects corrosion \cite{zhou2019simultaneous}, by passing a beam of protons through central part of a thin foil sample in contact with 650\textdegree C fluoride salt. Our experiments show that in all cases the unirradiated region suffers severe corrosion through the thickness of the foil, as evident by the penetration of salt to the other side. However, this barely occurs in the irradiated region. Figure 1a present a schematic of our molten salt corrosion experiments undergoing selected area proton irradiation, showing that the material incurs proton damage without hydrogen implantation. The beam-facing side of the samples (Figures 1e, 1h, and 1k) can be distinctly divided into two regions, whose boundaries match perfectly with the beam perimeter. Backscattered SEM images with corresponding elemental dispersive x-ray spectroscopy (EDX) point spectra reveal the cause of the color difference to be the existence of the salt along the grain boundaries on the outer (unirradiated) region of each sample. The region without irradiation (Figures 1f, 1i, and 1l) has salt decorating the grain boundaries, while the irradiated region (Figures 1d, 1g, and 1j) is almost free of salt. This distinct difference in the two regions remains when the beam current density is varied from 1.5-2.5~${\mu}$A/cm${^{2}}$. The proton irradiation, as the only difference between these two regions, is shown to be the reason for the slower molten salt penetration.  

To further confirm this effect, we ion-polished (Figure 2a) each sample to expose cross-sections and performed SEM characterization. Figures 2b-h show representative SEM images comparing the irradiated and unirradiated zones for different proton fluxes. In all cases, salt in the irradiated region rarely reaches the far side of the foil (Figures 2c-e), while salt readily penetrates through the foil in the $unirradiated$ region (Figures 2f-h). The corrosion is quite localized in our experiments such that the depth of corrosion varies significantly at different locations. Therefore, we have performed a statistical analysis to reveal the distribution of corrosion depth. Our analysis process (Figure 2i) starts from recording an ultra-high resolution image in each region, representing an area of 1 mm along the $y$-axis and 30$\mu$m along the $x$ axis. Afterwords, we divided the width of each image into more than 20,000 pixel rows, and calculated the maximum corrosion depth normalized by the sample thickness along each pixel row. To make quantitative comparisons, we plot the cumulative distribution function (CDF) of the normalized corrosion depth in Figure 2j. Comparing the the overall distributions of the irradiated and unirradiated regions at the same beam current, corrosion is clearly more severe in the regions without irradiation.

The wide distribution in corrosion depth is explained by both orientation-dependent attack and a mixture of simultaneous intergranular and transgranular corrosion. Here transgranular corrosion manifests as salt-filled voids penetrating transverse to grain boundaries. Once initiated, these voids can grow in depth following the GBs, at the same time expanding to one or both sides of the adjacent grains. Thus, a more representative and quantitative distinction between the two sample regions can be made by comparing the extent of deepest attacks for each region, in effect comparing GBs possessing the most susceptible characters without directly measuring their orientation relationships. We therefore selected the 10\% deepest attacks in each region (shown in the orange window in Figure 2j) and calculated their averaged normalized corrosion depth. As shown in Figure 2k, the irradiated region is corroded roughly two times less than the corresponding unirradiated region.

At this point there exist two possibilities to explain our observation of radiation-decelerated corrosion: (1) irradiation makes the alloy more corrosion-resistant; or (2) irradiation changes the chemistry of the molten salt so that it is less corrosive. To understand which one is valid, we have performed a comparative experiment on a pure Fe foil. Pure Fe corrodes much more uniformly in molten salts than a binary alloy because it contains no elements to selectively dissolve. Nonetheless the corrosion still results in surface asperities \cite{wang2016effects}. Corrosion etches valleys along GBs on the surface, which directly indicate the severity of the attack. We have collected a very large SEM composite image (26k by 26k pixels, 8 mm by 8 mm in size) of the salt-facing side of the pure Fe foil (Figure 3b). Zoomed-in images from the edge (Figure 3c) and the center (Figure 3d) show that the center suffers from more severe corrosion than the edge, implying that irradiation enhances corrosion in pure Fe in molten salts.

We then applied a machine-learning-based algorithm to partition the corroded and uncorroded regions (Figure 3e). The corroded region (shown in red) does not overlap with the beam profile (white circle) very well, suggesting that local variation of corrosion severity in this experiment is caused by the interaction between the proton beam and the molten salt rather than the Fe foil itself. Irradiation renders the salt to become more corrosive, and this modified salt can flow away from the irradiation region. As a result, we also see some local radiation-free regions experiencing more severe corrosion. Our experiments on pure Fe indicate that proton irradiation accelerates corrosion of pure elements via increasing the corrosiveness of the molten salt. Thus, we confirm that the interaction between irradiation and the Ni-Cr alloy is the dominant cause of radiation-decelerated corrosion. One should note that the phenomenon of irradiation increasing the corrosiveness of salt would also exist in the case of Ni-20Cr corrosion. However, it is outperformed by the radiation damage-decelerated corrosion in our experiments. As it scales up with increasing proton beam current, we would expect this effect to be more prominent at higher beam current. Indeed, we found that as the beam current is increased, the overall corrosion attack of Ni-20Cr becomes more severe in both irradiated and unirradiated regions, as shown in Figure 2k.

A model considering radiation-enhanced, bulk diffusion is proposed to explain how irradiation enhances corrosion resistance in Figures 4a-b. In our Ni-Cr binary alloy, Cr is preferentially depleted by molten salt because the redox potential of Cr is considerably lower than that of Ni in our LiF-NaF-KF (FLiNaK) fluoride salt \cite{zhang2018redox}. Cr in the bulk would diffuse out of the system via fast routes such as GBs to reach the salt \cite{dai2018corrosion}. The outward mass flux from GB to salt is compensated by the diffusion of lattice atoms (Ni and Cr) from the bulk to GBs, creating a self-healing mechanism to inhibit void formation along GBs. However, since bulk diffusion is much slower than GB diffusion at these temperatures, the vacancy density at GBs would increase during corrosion. As a result of corrosion, the atomic density of Cr at GBs would decrease, accompanied by an increase in Ni atomic density. Therefore, Ni is enriched and Cr is depleted along GBs (see supplementary figure S1). One should note that the bulk diffusion of Cr and Ni to GBs is just a "pain-reliever," but does not fully cease the increase of free volume at GBs. Eventually voids will nucleate at the GBs. Note that such bulk diffusion with or without irradiation is defined as: 
\begin{equation}
    D_{\text{total}} = D_{\text{i}} C_{\text{i}} + D_{\text{v}} C_{\text{v}}  \label{eq:diffisuion}
\end{equation}
where $D_i$ and $D_v$ represent diffusivities due to interstitial and vacancy mechanisms, respectively, and $C_i$ and $C_v$ are concentrations of interstitials and vacancies \cite{sizmann1978effect}. When there is no irradiation, the total diffusion flux of interstitials $\left(D_iC_i\right)$ is negligible due to the very low concentration of thermal interstitials, typically at least $10^6$ times lower than the corresponding vacancy flux \cite{was2016fundamentals}. However, the presence of irradiation will change the game.

Radiation damage cascades produce abundant interstitials within the grains in equal proportion to vacancies, which preferentially diffuse to defect sinks such as GBs \cite{bai2010efficient}. Therefore, irradiation activates the interstitial term in Equation \eqref{eq:diffisuion}, doubling the net flux of atoms towards GBs and thus accelerating the "self-healing" mechanism from its original, unirradiated rate. Because of that, void growth in grains adjacent to GBs will be much slower in the irradiated region than in the unirradiated region. In order words, irradiation enhances bulk diffusion and drives more atoms from the grain to the GBs to suppress void formation. Our model clarifies the mechanism of irradiation-decelerated intergranular corrosion to be enhanced mass transport to GBs via radiation-enhanced bulk diffusion. We emphasize that our model fundamentally differs from radiation induced segregation (RIS) \cite{ardell2016radiation}, in that corrosion implies an open system while RIS assumes a closed system. Our model is hypothesized to be effective in an alloy where one element is preferentially dissolved by a corrosive fluid. Therefore this effect should persist in other media, such as oxygen-poor molten lead, where selective dissolution is the dominant mode of corrosion \cite{zhang2009review}. Recent evidence even suggests that proton radiation can decelerate corrosion of stainless steels in high temperature water \cite{hanbury2019oxide}, challenging conventional wisdom even further. In a more general sense, radiation has also been noted to improve mechanical properties of structural materials in certain circumstances \cite{murty1984neutron}.

Last but not least, we compare the effects of proton and neutron irradiation on Ni-20Cr corrosion in molten salt. In our experiments, most protons stop in the salt, so the effect of injected hydrogen interstitials in the foil is negligible, though their effect on salt chemistry is strong. Unlike proton beams, neutrons will not introduce charged hydrogen when they stop in the salt. Thus, we expect that salt undergoing neutron irradiation may be less corrosive than that undergoing proton irradiation. As such, we expect that the irradiation-decelerated intergranular corrosion of Ni-Cr alloys under neutron irradiation to hold true, and perhaps even manifest a stronger deceleration compared to an equivalent flux of protons. It remains unclear whether the lower particle fluxes in nuclear reactors, compared to our experiments, will make this self-healing mechanism dominate over others. However, our results revealed an encouraging mechanism of irradiation slowing down intergranular corrosion via enhanced bulk diffusion, which has important implications for the rapid development and down-selection of structural materials to finally usher in advanced nuclear fission and fusion reactors.

\section*{Methods}
\subsection{Salt and material preparation}
The FLiNaK salt used for this work was produced in a temperature-controlled furnace housed inside an argon atmosphere glove box. Oxygen and moisture in the glove box were controlled to remain below 1~ppm. Powders of 99.99\% pure LiF (CAS number 7789-24-4), 99.99\% pure NaF (CAS number 7681-49-4), 99.99\% pure KF (CAS number 7789-23-3) and 99.98\% pure EuF\protect\textsubscript{3} (CAS number 13765-25-8) were purchased from Alfa Aesar. LiF, NaF, and KF powders were melted into separate discs at 1000$^{\circ}$C in glassy carbon crucibles supplied by HTW Germany. EuF\protect\textsubscript{3} powders were baked at 1000$^{\circ}$C. The salt discs of LiF, NaF, and KF were then broken into small pieces and measured by weight corresponding to the composition of a FLiNaK eutectic mixture (LiF-NaF-KF (46.5-11.5-42 mol~\%)). Baked EuF\protect\textsubscript{3} powder was then added corresponding to be 5~wt.~\% of the total. Addition of EuF\protect\textsubscript{3} increases the overall corrosion rate by increasing the redox potential of the salt \cite{guo2017measurement}. EuF\protect\textsubscript{3}, when reacting with Cr, becomes EuF\protect\textsubscript{2}, which leaves no deposition on the sample surface. Afterwards, the salt mixture was melted inside another glassy carbon crucible by holding at 700$^{\circ}$C for 12 hours. The final salt disc was then broken into small pieces. It was measured and melted into pellets roughly 3.5~g in weight. These pellets comprised the salt loaded into the corrosion/irradiation experiment.

An 80Ni-20Cr~wt.\% model alloy was made by Sophisticated Alloys Inc. in the form of 0.5~mm thick sheet with a certified purity level of 99.99\%. Then 30~\(\mu\)m thick foils of this alloy were rolled by the H. Cross Company. All samples studied in this work consisted of 14~mm diameter discs sectioned from the same rolled foil. Sample surfaces were smooth enough for the corrosion experiment without additional polishing.

For the pure Fe corrosion/irradiation experiment, a 99.5\% (metals basis) pure Fe foil was purchased from Alfa Aesar with a thickness of 25\(\mu\)m and used as-is.

\subsection{Simultaneous irradiation and corrosion experiments}
Details of the facility used in this study can be found in this reference \cite{zhou2019simultaneous}. Salt and sample loading were both performed in the same glove box mentioned previously. Then the vacuum-tight assembly was transferred out of the glove box to be connected to the proton accelerator beam line. The CLASS (Cambridge Laboratory for Accelerator Science) 1.7~MV Tandetron was used to accelerate protons to 3.0~MeV at the beam current densities listed in Figure 1. After the pressure reached \(10^{-6}\) Torr, the heater was started to reach 400$^{\circ}$C for at least half an hour, or until the pressure once again dropped below \(10^{-6}\) Torr, this ensured a proper bake-out of the as-assembled experimental facility. The temperature was increased to 650$^{\circ}$C over a period of roughly one hour. Then the proton beam with controlled energy, beam shape, and current was introduced to the sample by withdrawing the Faraday cup. The beam current during the experiment was controlled to be between \(\pm\)5\% of the target beam current. Temperatures of the control thermocouple were maintained at \(\pm\)1$^{\circ}$C. The two sides of the sample foil shared the same level of vacuum pressure during the experiments. Once the targeted duration of the experiment was reached, the beam was cut off, and the heater was stopped. It took less than half hour to cool down below 450$^{\circ}$C. After the system reached room temperatures, the assembly was backfilled with ultra-high purity argon. Then it was disconnected from the beam line, transferred into the glove box, and disassembled. 

\subsection{SEM analyses}
Prior to removal from the corrosion cell, pictures of the beam-facing side of the foils were taken by a camera, shown in Figures 1e, 1h, and 1k. Then the foils were cut out along their outer edge. The beam-facing side of each foil were imaged by a Phenom XL scanning electron microscope (SEM)in backscattering mode. These are shown in Figures 1d, 1f-g, 1i-j, and 1l. Cross sections were obtained by the method shown in Figure 2a using a JEOL SM-Z04004T argon ion polisher with a voltage of 6~kV and a beam current between 150 and 180~\(\mu\)A. A cross section wider than 1mm was obtained by polishing for over an hour. This was repeated for all Ni-20Cr samples, and representative SEM images of each are shown in Figures 2c-h.

\subsection{Cross-sectional image analysis of Ni-20Cr foils}

A grid of SEM images of each post-corrosion sample with the same magnification were taken in sequence, comprising an area 1~mm long. Grid stitching was performed using the stitching plugin \cite{preibisch2009globally} in Fiji \cite{schindelin2012fiji}. Then the stitched image was straightened and cropped corresponding to 1mm in width. Each image was then binarized with the auto threshold function of Fiji. For different sample foils, the contrast and brightness of the SEM images were slightly different. Thus, the auto threshold method was chosen so that the black pixels fully represented the salt-containing voids in the original SEM images. At the same time, patterns and artifacts from cross section polishing along the argon beam-facing side of the foil were contrast-enhanced to become black pixels. These artificial black pixels were removed by comparing to the original SEM images. Then, along each pixel row perpendicular to the sample surface, the black pixel farthest away from the salt-facing side was detected and marked as the corrosion depth for this pixel row. These data comprise the graph in Figure 2j. 

\subsection{Image analyses of the salt-facing side of the Fe foil}

After the corrosion/irradiation experiment, the pure Fe foil was soaked in deionized water for over 24 hours to remove the salt attached to its surface. Then the salt facing side of the Fe foil was imaged with the same SEM. 8mm by 8mm areas corresponding to 37 by 37 tiles of images were acquired. Then the 1369 tiles were stitched using the stitching plugin \cite{preibisch2009globally} in Fiji \cite{schindelin2012fiji}. As the corroded region exhibited preferentially etched grain boundaries while inside the grains remained clean, a simple color threshold method could not partition the the corroded and uncorroded regions for the Fe sample after molten salt corrosion. Therefore, a machine-learning based image segmentation method, called Trainable Weka Segmentation \cite{Arganda-Carreras2017}, was used. First, a small part of the image containing both corrded and uncorrroded regions was selected. Its size was 1/729$^{th}$ of the total area, and it was used to train a classifier model. Some of the regions were manually labeled as a training data set, the training and manual labeling process was iteratively performed several times until the classifier could correctly label all regions in the selection. Then the trained model was used to segment the entire high resolution image on a workstation with 512 GB RAM, which took about 8 hours.

\subsection{TEM analysis of corroded/irradiated grain boundaries}

TEM specimens were prepared by Ga$^+$ focused ion beam (FIB) lift-out from a cross-section of each foil. The beam energy used for thinning the sample was gradually reduced from 30 keV to 2 keV. STEM-EDX characterization was performed using an FEI ThemIS TEM at the national center for electron microscopy (NCEM) in the Molecular Foundry of Lawrence Berkeley National Laboratory. The TEM was operating in STEM mode with a electron beam energy of 300 keV. Linescans of Ni and Cr intensity were acquired across selected grain boundaries to obtain evidence of Ni enrichment and Cr depletion.

\section*{Acknowledgments}
The authors gratefully acknowledge funding from the Transatomic Power Corporation under Grant No. 023875-001, and the US Department of Energy Nuclear Energy University Program (NEUP) under Grant No. 327075-875J. A.M. acknowledges the support of FUTURE (Fundamental Understanding of Transport Under Reactor Extremes), an Energy Frontier Research Center funded by the U.S. Department of Energy, Office of Science, Basic Energy Sciences. Y.Y. was supported by the Director, Office of Science, Office of Basic Energy Sciences, Materials Sciences and Engineering Division, of the U.S. Department of Energy under Contract No. DE-AC02-05-CH11231 within the Mechanical Behavior of Materials (KC 13) program at the Lawrence Berkeley National Laboratory. The authors acknowledge support by the Molecular Foundry at Lawrence Berkeley National Laboratory, which is supported by the U.S. Department of Energy under Contract No. DE-AC02-05CH11231. The authors wish to thank Raluca Scarlat (UC Berkeley), Gabriel Meric de Bellefon (Kairos Power), En-Hou Han (IMR, China), and Il-Soon Hwang (UNIST, Korea) for discussions in guiding this study. Thanks are due to Mitchell Galanek, Ryan Toolin, Ed Lamere, and Ryan Samz from MIT's Environmental Health and Safety (EHS) department for verifying device safety and shielding, and to Amy Tatem-Bannister and William DiNatale for training and access to the argon ion cross section polisher.

\section*{Author Contributions}

W.Y.Z. constructed the corrosion/irradiation facility with assistance from G.Q.Z. K.B.W., and P.W.S. W.Y.Z. conducted all corrosion and irradiation experiments and SEM characterizations. Y.Y. and A.M. performed TEM sample preparation and TEM characterization as well as machine learning auto-identification of corroded regions on the pure Fe sample. M.P.S. conceived of the original project and oversaw its execution, providing regular guidance.  W.Y.Z, Y.Y and M.P.S wrote the manuscript. All authors contributed to the analysis of the results.

\section*{Competing Interests}

The authors declare no competing interests.

\section*{Materials \& Correspondence}

Please address all inquiries to Michael Short at hereiam@mit.edu

\section*{Data Availability}

All the data from this study, including original micrographs, processed data, scripts, machine learinng codes, and compiled results are are available from the corresponding authors upon reasonable request.

\bibliography{references}

\begin{figure*}[!t]
\begin{centering}
\includegraphics[width=1\linewidth]{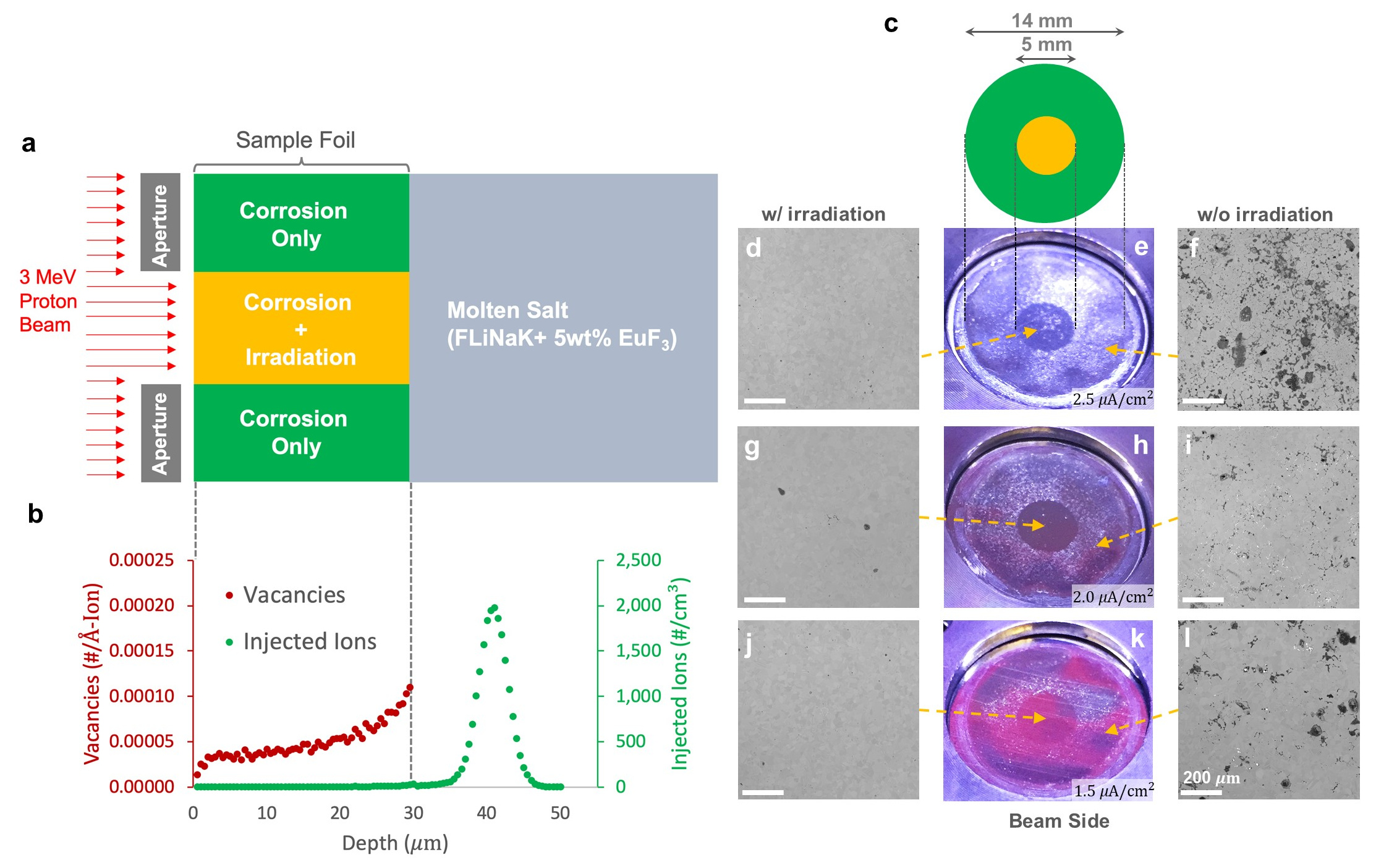}
\par\end{centering}
\caption{Schematics of the experimental setup, sample configuration, and beam-facing side comparison regions of Ni-20Cr samples. \textbf{(a)} The experimental setup, showing how the two sample regions were created. \textbf{(b)} Distribution of primary radiation damage (in red) and proton deposition (in green) along the irradiation direction for the Ni-20Cr corrosion experiments, simulated by SRIM \cite{ziegler2010srim}. \textbf{(c)} Schematic of the irradiated (green) and unirradiated (yellow) zones. \textbf{(d-l)} Optical and representative SEM images of the beam-facing side of the Ni-20Cr foils after 4 hours at 650$^{\circ}$C under various beam currents. (d-f), (g-i), and (j-l) correspond to beam currents of 2.5 ${\mu}$A/cm${^{2}}$, 2.0 ${\mu}$A/cm${^{2}}$, and 1.5 ${\mu}$A/cm${^{2}}$, respectively.\label{Figure 1}}
\end{figure*}

\begin{figure*}
\begin{centering}
\includegraphics[width=1\linewidth]{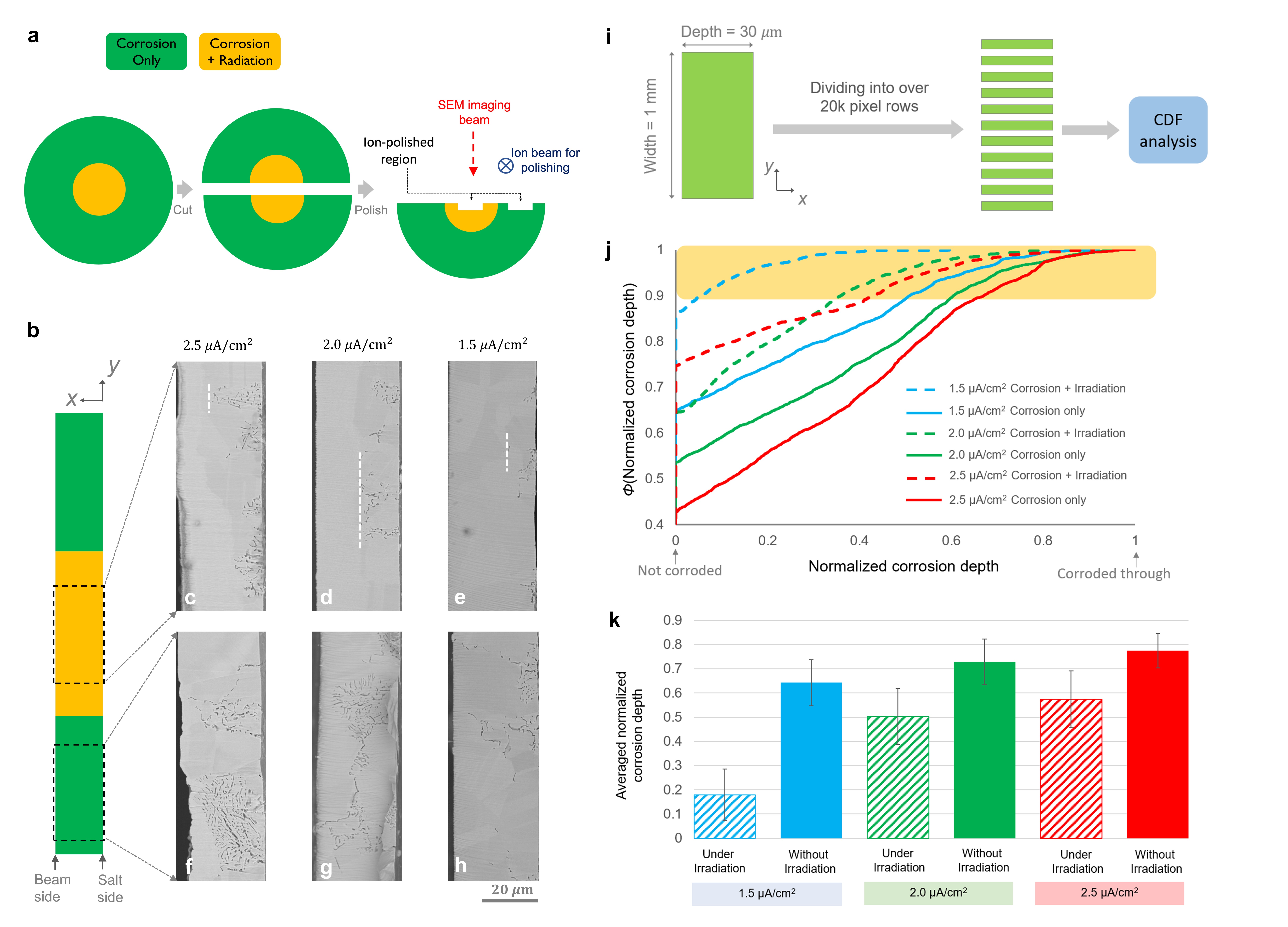}
\par\end{centering}
\caption {Cross-sectional comparison of corrosion with and without molten salt irradiation on Ni-20Cr. \textbf{(a)} Ion-polishing process used to prepare samples for cross-sectional SEM imaging. \textbf{(b-h)} Representative SEM images from different zones of Ni-20Cr samples under different beam currents. \textbf{(c-e)} show the irradiated zone, with white dashed-lines indicating the deepest attack. \textbf{(i)} Data analysis process for the Ni-20Cr foils. \textbf{(j)} Cumulative distribution function (CDF) of corrosion depth normalized by sample thickness with data from Figure (i). \textbf{(k)} Average, normalized corrosion depth using data from the 10\% deepest corroded regions of each sample, as illustrated by the orange bar in (j). \label{Figure 2} }
\end{figure*}

\begin{figure*}
\begin{centering}
\includegraphics[width=1\linewidth]{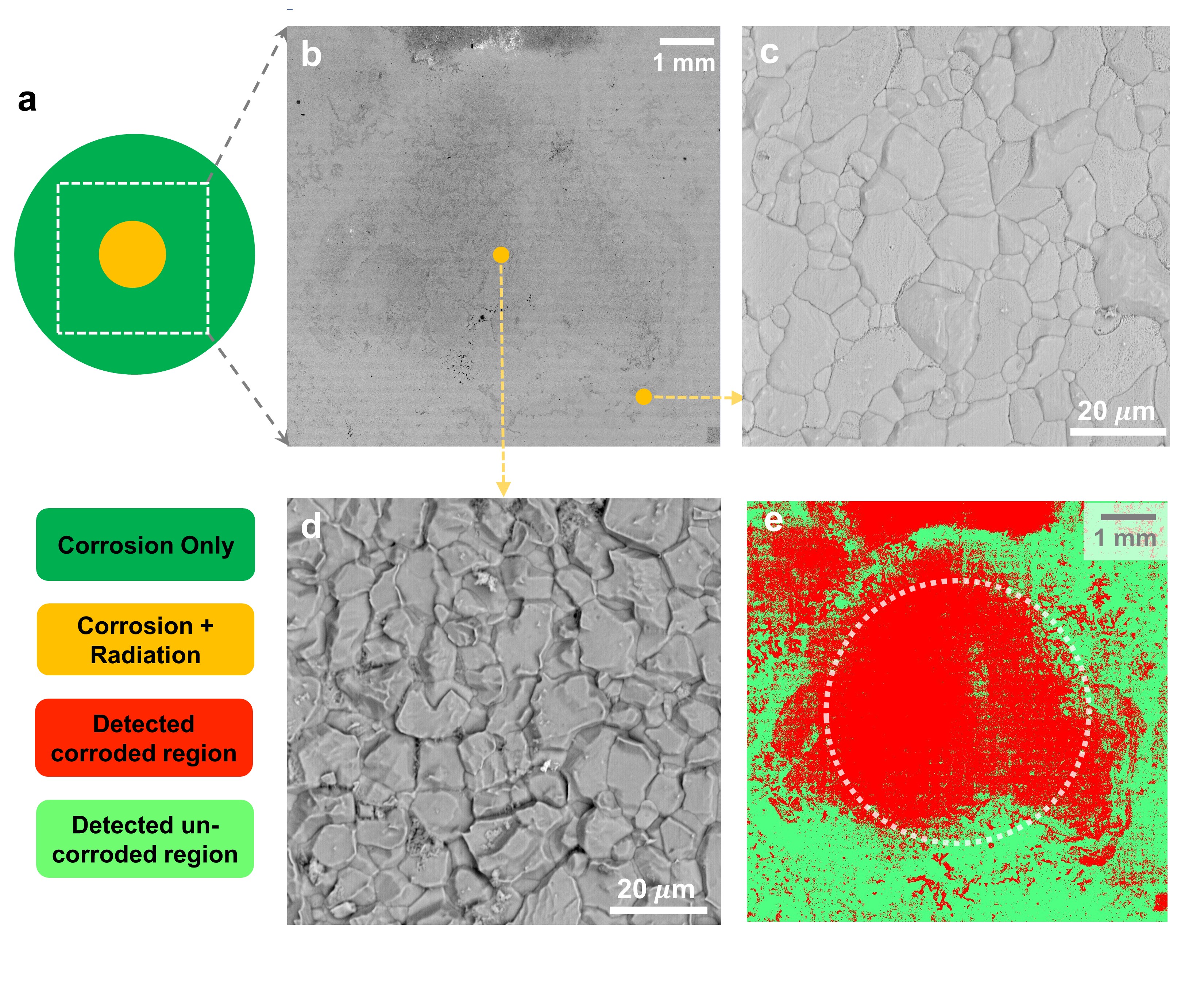}
\par\end{centering}
\caption{Salt-facing side of pure Fe following simultaneous corrosion/irradiation. \textbf{(a)} Location on the Fe foil used for SEM imaging. \textbf{(b)} SEM image of the salt-facing side after 6~hours at 650$^{\circ}$C under 2.0~${\mu}$A/cm${^{2}}$ proton irradiation. \textbf{(c)} Enlarged SEM image for the edge (unirradiated) region of the foil. \textbf{(d)} Enlarged SEM image for the center (irradiated) region of the foil. \textbf{(e)} Machine-learning based segmentation of (b), showing auto-identified corroded and uncorroded regions. The white circle delineates the proton beam perimeter.
\label{Figure 3} }
\end{figure*}

\begin{figure*}
\begin{centering}
\includegraphics[width=1\linewidth]{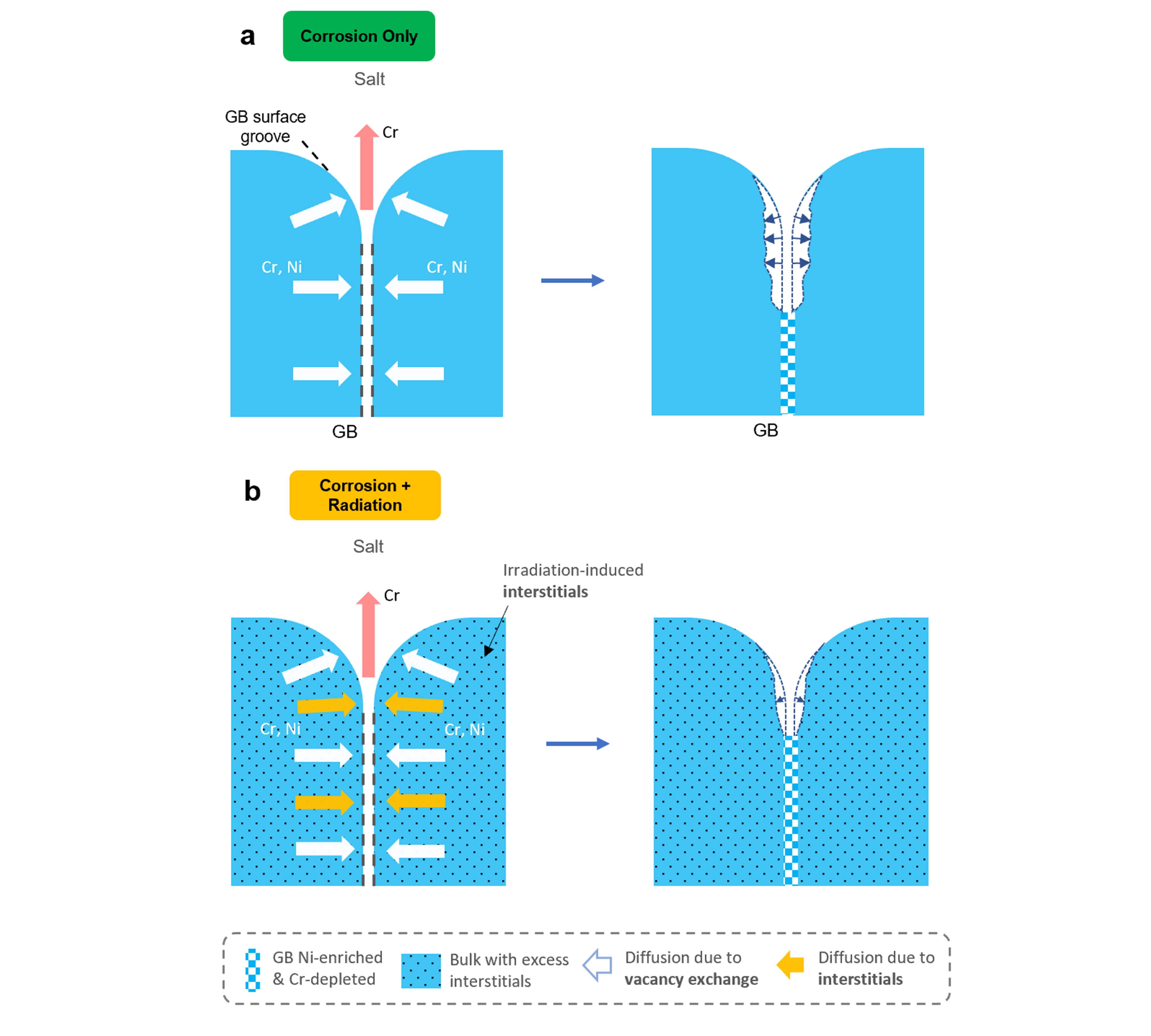}
\par\end{centering}
\caption{Proposed mechanism of radiation-decelerated corrosion. \textbf{(a)} Schematic of the solid state diffusion processes during molten salt corrosion in Ni-20Cr. \textbf{(b)} Schematic of the solid state diffusion processes during molten salt corrosion in Ni-20Cr under the influence of proton irradiation.\label{Figure 4}}
\end{figure*}

\renewcommand{\thefigure}{S\arabic{figure}}
\setcounter{figure}{0}
\begin{figure*}
\begin{centering}
\includegraphics[width=0.5\linewidth]{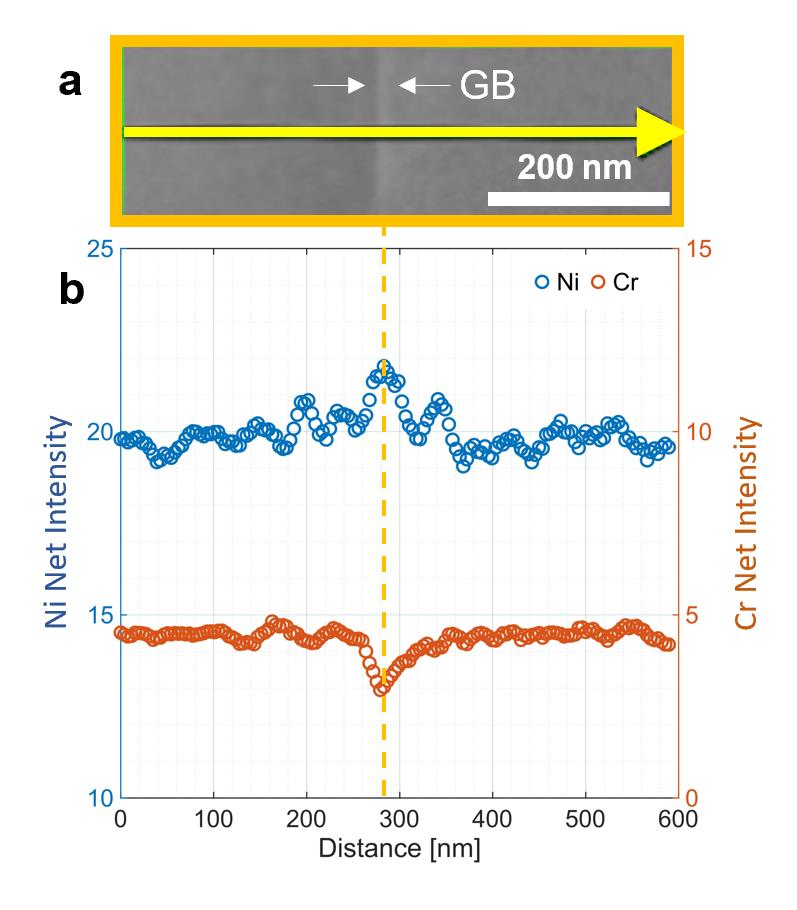}
\par\end{centering}
\caption{Elemental distribution results from TEM/EDX measurements. \textbf{(a)} STEM image of a characteristic grain boundary in the irradiated zone, as indicated by the white arrow. The yellow line marks the EDX linescan. \textbf{(b)} EDX net intensity profile for the line shown in (a), indicating that the atomic density of Ni is increased at the GB. \label{Figure S1}}
\end{figure*}

\end{document}